\documentstyle[12pt]{article}
\begin{document}
\title{Low energy wave packet tunneling from a parabolic potential well
through a high potential barrier}
\author{V. V. Dodonov$^{1,2,3}$, A. B. Klimov$^4$ and V.I. Man'ko$^{2,5}$\\
{\normalsize $^1$ Departamento de F\'{\i}sica,
Universidade Federal de S\~ao Carlos,}\\
{\normalsize Via Washington Luiz, km 235, S\~ao Carlos, SP, Brasil}\\
{\normalsize $^2$ Lebedev Physics Institute, Leninsky Prospect 53, 117924
Moscow, Russia}\\
{\normalsize $^3$ Moscow Institute of Physics and Technology}\\
{\normalsize $^4$ Departamento de F\'{\i}sica, Universidad de Guadalajara,}\\
{\normalsize Corregidora 500, 44420 Guadalajara, Jalisco, M\'exico}\\
{\normalsize $^5$       Oservatorio Astronomico di Capodimonte,}\\
{\normalsize       Via Moiariello, 16--80131, Napoli, Italy } }
\date{}
 \maketitle
\begin{abstract}
The problem of wave packet tunneling in a potential
$V(x)=(m\omega^2/2)\left(x^2 -\delta x^{\nu}\right)$ with $\nu>2$ is
considered in the case when the barrier height is much greater than
$\hbar\omega$ and the difference between the average energy of the packet
and the oscillator ground state energy $\hbar\omega/2$ is sufficiently small.
The universal Poisson distribution of the partial tunneling rates from
the oscillator energy levels is discovered. The explicit expressions for
the tunneling rates of different types of packets
(coherent, squeezed, even/odd, thermal, etc.)
are given in terms of the exponential and modified Bessel
functions. The tunneling rates turn out very sensitive to the energy
distributions in the packets, and they may exceed significantly the
tunneling rate from the energy state with the same average number of quanta.
\end{abstract}


\newpage \section{Introduction}
Usually, the problem of quantum tunneling through potential barriers was
considered under the assumption that the initial state possessed a well
defined energy, i.e., for the quasistationary states.
The propagation of the Gaussian {\it wave packets\/} through a rectangular
barrier was studied, e.g., in \cite{Hart}. Recently, the problem of decay of
{\it coherent\/} and {\it squeezed\/} packets confined initially in a deep
potential well of a finite depth was considered by several authors
\cite{Dek,Mug,Lenef}. The contribution of dissipation was studied, e.g., in
\cite{Cald,Raz,Mug-dis,Likh-book,Kam-Raz}. However, in all those papers
the influence of the wave packet shape on the transition
or escape rates was analyzed in the framework of numerical calculations only.

The aim of the present article is to present simple analytical formulas
for the decay rate in the case when a wave packet is localized
initially near to the bottom of a deep potential well
\begin{equation}V(x)=\frac 12 m\omega^2\left(x^2-\delta x^{\nu}\right),
\qquad \nu>2. \label{V(x)}\end{equation}
The special cases of this potential were considered in
\cite{Lenef,Cald,Likh-book,Kam-Raz,Likh} ($\nu=3$) and \cite{Langer} ($\nu=4$).
More precisely, we assume that the potential is close
to the harmonic one for relatively small values of $|x|$, while it goes to
$+\infty$ when $x\to-\infty$, so that we have a single barrier at $x>0$.
Besides, it is implied that the potential is given by Eq. (\ref{V(x)})
provided $x^2-\delta x^{\nu}>0$, whereas at large values of $x$
(to the right of the barrier) it tends to some constant value.

If the initial state were described by means of the {\it diagonal density
matrix\/} $ \hat{\rho}=\sum \rho_n |n\rangle\langle n|$,
then the total decay rate $\gamma$ would be a sum of partial rates
$\gamma_n$ taken with proper weights \cite{Likh-book,Likh,Aff,Goldan},
\begin{equation}\gamma=\sum_n\rho_n\gamma_n. \label{main}\end{equation}
For {\it pure\/} superpositions of many wave functions with different
energies $E_n$ the situation, in principle, may be more complicated due to
the possibility of quantum interference effects. However, under certain
conditions Eq. (\ref{main}) can be applied to the pure states, as well.
Suppose that we have a single high (unidimensional) barrier from the right,
so that the motion can be considered as free at $x>L$. Let us designate
the probability of discovering the particle in the well and under the barrier
as $w_L=\int_{-\infty}^L|\psi (x)|^2\,dx$ . Then an immediate consequence of
the quantum continuity equation is the relation $\dot {w}_L=-j(L)$, where
$ j=(i\hbar/2m)\left(\psi\nabla\psi^{*}-\psi^{*}\nabla\psi\right)$ is the
usual current density. If the wave function of the packet has the form
\begin{equation} \psi =\sum_n c_n\psi_ne^{-iE_nt/\hbar}\label{psi-sup}
\end{equation}
(it holds for decaying states, as well, provided that $t\ll\gamma^{-1}$), then
\[j=\sum_n|c_n|^2j_n+\sum_{m\neq n}j_{mn}e^{i(E_n-E_m)t/\hbar}.\]
Since the second sum consists of a large number of rapidly oscillating terms
with different phases and frequencies, it turns practically into zero for
$|E_n-E_m|t/\hbar\gg 1$, and we get Eq. (\ref{main}) with $\rho_n=|c_n|^2$,
$j_n=\gamma_n$, and $\gamma=\bar{j }$, the overbar meaning the averaging
over fast oscillations. This result holds under the condition
$ \hbar\gamma\ll \left|E_n -E_m\right|$, where suffices $n,m$ correspond to all
the coefficients $c_n$ that yield significant contributions to the expansion
(\ref{psi-sup}). In the special case of potential (\ref{V(x)}) with $\nu=3$,
Eq. (\ref{main}) was actually derived in \cite{Lenef} in the framework of the
quasiclassical method proposed in \cite{Dek}.
In the present paper we pursue two main goals: i) to find an analytical
expression for the partial decay rate $\gamma_n$ in the potential (\ref{V(x)}),
ii) to calculate the sum (\ref{main})
explicitly for different physically interesting initial wave packets.

\section{Partial decay rates from a parabolic well}
To calculate the partial tunneling rates we use the standard quasiclassical
formula \cite{Likh-book,Goldan}
\begin{equation} \gamma_n =\frac{\omega}{2\pi}\exp\left\{-\frac{2}{\hbar}
\sqrt{2m}\int_{x_1}^{x_2}\sqrt{V(x)-E_n}\,dx\right\}.\label{tunn}\end{equation}
It is convenient to rewrite it in the form
\begin{equation} \gamma_n^{(\nu)} =\frac{\omega}{2\pi}\exp\left\{-\frac{2V_0}
{\lambda_{\nu}\hbar\omega}F_{\nu}\left(2\lambda_{\nu}\frac{E_n}{V_0}\right)
\right\}, \label{tunnF}\end{equation}
where $V_0$ is the barrier height:
\[ V_0=m\omega^2\lambda_{\nu}\delta^{-2/(\nu-2)}, \qquad
\lambda_{\nu}=\frac{\nu-2}{2\nu}\left(\frac{2}{\nu}\right)^{2/(\nu-2)}.\]
Function $F_{\nu}(t)$ is given by the integral
\begin{equation} F_{\nu}(t) = \int^{b}_{a}dx\left[\matrix{x^{2}-x^{\nu}-t}
\right]^{1/2}, \label{F(t)}\end{equation}
where $a < b$ are positive solutions to the equation
\begin{equation}  x^{2}-x^{\nu}-t = 0. \label{eq-t}\end{equation}

It is known that for $\nu=3$ and $\nu=4$ the integral (\ref{F(t)}) can be
expressed in terms of the complete elliptic integrals of the first and
second kind \cite{Grad}
\[{\bf K}(z)=\int_0^1\frac {dx}{\sqrt {\left(1-x^2\right)\left(1-z^2x^2
\right)}}, \qquad {\bf E}(z)=\int_0^1\sqrt {\frac {1-z^2x^2}{1-x^2}}\,dx.\]
At $\nu=4$ the roots of Eq. (\ref{eq-t}) can be
found explicitly, and formula (3.155.9) of \cite{Grad} yields
\begin{equation} F_4(t)=\frac 13\left(1+\xi^2\right)^{-3/2}\left[\left(1+
\xi^2\right){\bf E}\left(\sqrt {1-\xi^2}\right)-2\xi^2{\bf K}\left(\sqrt
{1-\xi^2}\right)\right], \label{F4}\end{equation}
\[\xi^2=\frac{1-\sqrt{1-4t}}{1+\sqrt{1-4t}}.\]
At $\nu=3$ we get
\begin{equation}
F_3(t)=\frac 2{15}\left(1-\xi^2+\xi^4\right)^{-5/4}\left\{2\left(1-\xi^2+
\xi^4\right){\bf E}\left(\sqrt {1-\xi^2}\right)-\xi^2\left(1+\xi^2\right)
{\bf K}\left(\sqrt {1-\xi^2}\right)\right\},\label{F3}\end{equation}
where $\xi^2=(a-c)/(b-c)$, and $c$ is the negative root of Eq. (\ref{eq-t}).
Another expression for $\gamma_n^{(3)}$ was given in \cite{Lenef}.

Assuming that coefficient $\delta$ in the potential energy (\ref{V(x)})
is sufficiently small, we have $Q\equiv V_{0}/(\hbar \omega ) \gg 1 $.
Then $m\omega x_*^2/\hbar=2\nu Q/(\nu-2)\gg 1$, where $x_*$ is the position
of the maximum of the potential energy. This means that the energies of the
low levels practically coincide with the harmonic oscillator energy
$E_n=\hbar \omega (n+{1\over 2})$. For these levels $E_n\ll V_0$, so we need
the expansions of the exact expressions (\ref{F4}) and (\ref{F3})
at $t\ll 1$. For $\nu=4$, the known asymptotics of the complete elliptic
integrals \cite{Grad} results in the formula
\begin{equation} F_4(t)=\frac 13 +\frac t4\ln\left(\frac t{16e}\right)+
\frac 3{32}t^2\ln t +{\cal O}\left(t^2\right). \label{F4as}\end{equation}
For $\nu=3$ the roots of the cubic equation (\ref{eq-t}) read
(to within an accuracy of the order of $t^{3/2}$)
$$ a =\sqrt{t}\left(1+\frac{\sqrt{t}}{2}\right),\quad
c = -\sqrt{t}\left(1-\frac{\sqrt{t}}{2}\right),\quad  b = 1-t,\quad
\xi ^{2} = 2\sqrt{t}(1-\sqrt{t}).$$
However, the expansion of $F_3(t)$ does not contain odd powers of $\sqrt{t}$:
\begin{equation} F_3(t)=\frac 4{15}+\frac t4\ln\left(\frac t{64e}\right)+
{\cal O}\left(t^2\ln t\right). \label{F3as}\end{equation}

Both the expressions, (\ref{F4as}) and (\ref{F3as}), contain the same term
$(t/4)\ln(t/e)$. This coincidence is not accidental: the leading term in
the expansion of the integral (\ref{F(t)}) at $t\to 0$ equals $(t/4)\ln(t/e)$
{\it for any\/} $\nu >2$. Indeed, the contribution to this integral of the
domain near the left turning point can be represented as
\[F_{left}=\int_{\sqrt{t}}^A \sqrt{x^2-t}\,dx +\cdots\,,\]
where $A$ is some finite number. Thus $F_{left}=(t/4)\sinh(2z)-tz/2 +\cdots$,
where $A=\sqrt{t}\cosh(z)$. In the limit of $A^2/t\gg 1$ we get
\[F_{left}=A^2/2+(t/4)(\ln{t}-1) -(t/2)\ln(2A)+{\cal O}(t^2/A^2).\]
In the vicinity of the right turning point we put $x=1-\varepsilon$ and write
\[F_{right}=\int_{t/(\nu-2)}^B\sqrt{(\nu-2)\varepsilon-t}\,d\varepsilon
+\cdots =\frac23 B^{3/2}\sqrt{\nu-2}-t\sqrt{\frac{B}{\nu-2}}+{\cal O}(t^2).\]
As to the integral in the limits from $A$ to $B$, it has an obvious power
expansion with respect to $t$. Consequently, the following expansion holds:
\[F_{\nu}(t)=f_{\nu}^{(0)}+ (t/4)\ln(t/e)-f_{\nu}^{(1)}t +{\cal O}\left
(t^2\ln t\right),\]
where coefficients $f_{\nu}^{(0)}$ and $f_{\nu}^{(1)}$ depend on the concrete
value of the exponent $\nu$. Then the partial tunneling rate reads
\begin{equation}\gamma_n^{(\nu)}=\frac {\omega}{2\pi}\exp\left\{-\frac {2Q}
{\lambda_{\nu}}f_{\nu}^{(0)} + (n+ 1/2)\left[\ln\left(\frac {eQ}{2(n+ 1/2)
\lambda_{\nu}}\right)+ 4f_{\nu}^{(1)}\right] +{\cal O}\left(n^2\frac {\ln Q}
{ Q}\right)\right\}. \label{tau-nQ}\end{equation}

Now let us notice that the factor $(n+ 1/2)[\ln(n+ 1/2)-1]$ is the leading
term of Stirling's asymptotical formula
\[\ln(n!)=(n+ 1/2)[\ln(n+ 1/2)-1]+(1/2)\ln(2\pi)+{\cal O}(1/n),\]
which works quite well even at $n\approx 1$.
Consequently, under the restriction
\begin{equation} n^{2}\ln (Q)/ Q\ll 1 \label{restr}\end{equation}
the partial decay rates are given by the Poisson distribution:
\begin{equation} \gamma_n^{(\nu)}=\gamma_0^{(\nu)}\frac{\chi_{\nu}^n}{n!},
\label{tau-Pois}\end{equation}
\[\chi_{\nu}=\mu_{\nu}Q,\qquad \mu_{\nu}=(2\lambda_{\nu})^{-1}\exp\left[4f_
{\nu}^{(1)}\right],\qquad \gamma_0^{(\nu)}=\omega{}\sqrt{\frac{\chi_{\nu}}
{2\pi}}\exp\left[-\frac{2 Q}{\lambda_{\nu}}f_{\nu}^{(0)}\right].\]
Specifically,
\[\mu_3=432,\qquad \mu_4=64, \qquad \gamma_0^{(3)}=\omega\sqrt{\frac{216}
{\pi}Q}\exp\left(-\frac{36}5 Q\right), \qquad
\gamma_0^{(4)}=\omega\sqrt{\frac{32}{\pi}Q}\exp\left(-\frac{16}3 Q\right).\]

Strictly speaking, the right-hand side of Eq. (\ref{tunn}) contains some
additional preexponential factor $G\left(V_0,\omega,E_n\right)$. But this
factor is a smooth function of energy. This means that $G\approx G_0\left[
1+{\cal O}(n/Q)\right]$, while the leading exponential term was approximated
with an accuracy of the order of $n^{2}\ln{Q}/Q$. Consequently, the influence
of the preexponential factor can be neglected under the restriction
(\ref{restr}). Note that our expression for $\gamma_0^{(3)}$ coincides
identically with the result of \cite{Likh-book}, where a special attention
was paid to the correct calculation of the preexponential term.

Formula (\ref{tau-Pois}) seems to be a universal distribution of the partial
decay rates from the low energy levels, which holds for any potential
of the form $V(x)=(m\omega^2/2)[x^2-u(x)]$ with $|u(x)|\ll x^2$ at $x\to 0$,
provided that conditions $Q\gg 1$ and (\ref{restr}) are fulfilled.
The concrete form of $u(x)$ is responsible for the precise value of the
coefficient $\mu$. Since $\chi_{\nu}\gg 1$, the total decay rate turns out
very sensitive to the detailes of the distribution $\rho_n$.

\section{Decay of a slightly deformed ground state}
The simplest example of the initial wave packet corresponds to a
{\it coherent\/} state, i.e., an eigenstate of the operator
$\hat{a}=\left(m\omega \hat{x}+i\hat{p}\right)/\sqrt{2m\omega\hbar}$. Then
\begin{equation}\rho_n(\alpha )=\frac {|\alpha |^{2n}}{n!}e^{-|\alpha|^2},
\qquad |\alpha|^2=\bar{n},\label{dist-coh}\end{equation}
and Eqs. (\ref{main}), (\ref{tau-Pois}) and (\ref{dist-coh})
result in the formula (we drop the suffix $\nu$)
\begin{equation} \gamma_{coh}=\gamma_0 \exp\left(-|\alpha|^2\right)
I_0\left(2|\alpha|\sqrt{\chi}\right)=\gamma_0 \exp\left({-\bar{n}}\right)
I_0\left(2\sqrt{\chi\bar{n}}\right)  , \label{gam-coh}\end{equation}
$I_0(z)$ being the modified Bessel function. To evaluate the domain of its
validity, we notice that the maximal contribution to the sum (\ref{main}) is
given by the terms with $n_{max}\sim \sqrt{\chi\bar{n}}$, and the width
of the effective distribution $\gamma_n\rho_n$ is obviously less than
$\sqrt{n_{max}}$. Thus the requirement (\ref{restr}) results in the inequality
$|\alpha|^2=\bar{n}\ll 1/\ln{Q}$, so that the term $\exp\left({-\bar{n}}
\right)$ can be omitted. However, the condition $\bar{n}\ll 1/\ln{Q}$ does
not exclude the possibility of $\sqrt{\chi\bar{n}}\gg 1$. In this case we have
\begin{equation} \gamma_{coh} =\gamma_0\left(4\pi\sqrt{\chi\bar{n}}\right)
^{-1/2}\exp\left(2\sqrt{\chi\bar{n}}\right)\gg\gamma_0.
\label{gam-cohas}\end{equation}

The most general {\it pure\/} Gaussian state (which is called frequently
in the current literature as a ``squeezed state'': see \cite{200} and
references therein) can be considered as an eigenstate of a linear
combination of the operators $\hat {a}$ and $\hat {a}^{\dag}$:
\begin{equation} \hat {b}|\beta uv\rangle=\beta|\beta uv\rangle, \qquad
\hat {b}=u\hat {a}+v\hat {a}^{\dag}, \qquad |u|^2-|v|^2=1
\label{b-uv}\end{equation}
(for simplicity, we confine ourselves to the case of linear {\it uniform\/}
transformations). The corresponding level population distribution
reads \cite{200,Yuen}
\begin{equation}\rho_n=|\left\langle n|\beta uv\right\rangle |^2=\frac
1{|u|n!}\left|\frac v{2u}\right|^n\exp\left[-|\beta |^2+\mbox{Re}\left
(\beta^2\frac {v^{*}}u\right)\right]\left|H_n\left(\frac {\beta}{\sqrt {
2uv}}\right)\right|^2,\label{200447}\end{equation}
where $H_n(x)$ is the Hermite polynomial. To calculate the sum (\ref{main}) we
need a formula for $\sum {g^n}H_n(x)H_n(x^*)/(n!)^2$. It can be easily found,
if one takes the known generating function of the Hermite polynomials
\[\sum_{n=0}^{\infty}\frac {z^n}{n!}H_n(x)=\exp\left(2xz-z^2\right),\]
multiplies both sides by the complex conjugated functions, puts $z=\sqrt{g}
\exp(i\varphi)$, and integrates the product over $\varphi$. The result is
\[\sum_{n=0}^{\infty}\frac {g^n}{(n!)^2}\left|H_n(x)\right|^2=
\frac{1}{2\pi}\int_0^{2\pi}\exp\left[4|x|\sqrt{g}\cos(\varphi+\psi)-
2g\cos(2\varphi)\right]\,d\varphi,\qquad x=|x|e^{i\psi}.\]
In this way we get the expression
\begin{equation} \gamma_{sq} = \frac{\gamma_0}{2\pi|u|}\int_0^{2\pi}\exp
\left[2\left|\frac{\beta}{u}\right|\sqrt{\chi}\cos(\varphi+\psi)
- \chi\left|\frac{v}{u}\right|\cos(2\varphi)\right]\,d\varphi,\qquad
\psi=\arg\left(\frac{\beta}{\sqrt{uv}}\right) \label{gam-squeez}\end{equation}
(we neglect the contribution of the terms proportional to $|\beta|^2$
in the argument of the exponential function, since it is very small under
the conditions $\chi\gg 1$ and $|\beta|\ll 1$, which ensure the validity
of Eq. (\ref{gam-squeez})). At $v=0$ we arrive again at Eq. (\ref{gam-coh}).

For a {\it slightly squeezed\/} state with $\chi|v|\ll |\beta|\sqrt{\chi}$
(this inequality implies $|v|\ll 1$, so that $|u|$ can be replaced by unity)
the integral in (\ref{gam-squeez}) can be calculated with the aid of the
steepest descent method, provided that $|\beta|\sqrt{\chi}\gg 1$. The
integrand assumes its maximal value at $\varphi=-\psi$. Thus we get
\[\gamma_{sq}(\beta,v) =\gamma_{coh}(\beta)\exp\left[
- \chi|v|\cos(2\psi)\right].\]
Consequently, the decay rate of a squeezed packet may be both greater and
less than the decay rate of the coherent packet with the same mean energy,
depending on the value of the phase difference $\psi$. (This qualitative
result was obtained in \cite{200,Sing} for $\nu=3$, although the quantitative
estimations were not quite correct, since the importance of logarithmic
terms in the expansion of $F_{3}(t)$ was underestimated.)
To elucidate the situation, we take into account the formulas for the
average number of quanta and its variance in the squeezed state \cite{200}
\[\bar {n}=|v|^2\left(1+|\beta |^2\right)+|u|^2\left[|\beta |^2-2\mbox{Re}
\left(\beta^2\frac {v^{*}}u\right)\right],\]
\[\sigma_n\equiv\overline {n^2}-(\bar {n})^2=2\bar {n}\left[2|u|^2
-1\right]-2|v|^4-|\beta |^2.\]
They can be simplified significantly if $|v|\ll 1, |u|\approx 1$:
\[\bar {n}\approx|\beta|^2[1-2|v|\cos(2\psi)],\qquad
\sigma_n\approx 2\bar {n}-|\beta|^2.\]
Finally, we get
\begin{equation} \gamma_{sq}=\gamma_0\left[(4\pi)^2\chi\bar {n}\right]^{-1/4}
\exp\left[2\sqrt{\chi\bar {n}}+\frac12\chi S\right],\qquad
S=\left(\sigma_n -\bar{n}\right)/\bar{n}, \label{rate-sq}\end{equation}
where $S$ is the known Mandel's parameter characterizing the type of
photon statistics. In this case the super-Poissonian statistics enhances
the tunneling rate, while the sub-Poissonian one suppresses it.
A different dependence of the tunneling rate on the degree of squeezing
was found in Ref. \cite{Lenef}. But its authors performed the numerical
calculations in a quite different domain of parameters: $Q=10$ and
$|\beta|\ge 1$, where our approach cannot be applied.

In the case of a {\it squeezed vacuum\/} ($\beta=0$) the right-hand side
of Eq. (\ref{gam-squeez}) coincides with the known integral representation
of the modified Bessel function. Then
\begin{equation} \gamma_v=\gamma_0 I_0(\chi|v|)=\gamma_0 I_0(\chi
\sqrt{\bar{n}}), \label{sq-vac}\end{equation}
provided that $|v|^2=\bar{n}\ll( Q\ln Q)^{-1}$ (for this reason we
put $|u|=1$). If $\chi|v|\gg 1$ and $|\beta|\ll|v|\sqrt{\chi}$, then one
can use again the steepest descent method. Now we have two extremal points:
$\varphi=\pi/2$ and $\varphi=3\pi/2$, so
\begin{equation} \gamma_v(\beta)=\gamma_0\left(2\pi\chi|v|\right)^{-1/2}
\exp(\chi|v|)\cosh(2|\beta|\sqrt{\chi}\sin\psi). \label{sq-vbet}\end{equation}
In this case we have $\gamma_v(\beta)>\gamma_v$, although Mandel's parameter
$$ S=\left[|v|^2-2|\beta|^2|v|\cos(2\psi)\right]
/\left(|v|^2 + |\beta|^2\right)$$
may be both positive and negative, in spite of the requirement
$|\beta|\ll|v|\sqrt{\chi}$.

The level populations in a Gaussian {\it mixed\/} state with {\it zero\/} mean
values of the quadratures are expressed in terms of the Legendre polynomials
\cite{Sem,Sriniv,DOM}:
\begin{equation}\rho_n=2(4d+ 2T+ 1)^{-1/2}\left(\frac{4d+ 1- 2T}{4d+ 1+ 2T}
\right)^{n/2}P_n\left(\frac {4d- 1}{\left[\left(4d+ 1\right)^2-
4T^2\right]^{1/2}}\right).\label{dist-mix0}\end{equation}
Parameters $d$ and $T$ are related to the ``degree of mixing'' of the
quantum state and the mean quantum number:
\begin{equation} d^{-1}=4\left[\mbox{Tr}\left(\hat\rho^2\right)\right]^2,
\qquad T=1+2\bar{n}, \qquad T\ge 2\sqrt{d}. \label{d-T}\end{equation}
In this case the sum in the right-hand side of Eq. (\ref{main}) is reduced to
the known generating function of the Legendre polynomials (see, e.g.,
Eq. 10.10(40) from \cite{Bateman}). Since we are restricted with the inequality
$\bar{n}\ll 1$, it is convenient to introduce a small parameter
$\varepsilon\le\bar{n}\ll 1$ according to the relations
$\mbox{Tr}\left(\hat\rho^2\right)\approx 1-2\varepsilon$,
$d\approx 1/4 + \varepsilon$. Then we get a simple formula
\begin{equation}
\gamma_{gauss}(\chi,\bar{n},\varepsilon)=\gamma_0\exp(\chi\varepsilon)
I_0\left(\chi\sqrt{\bar{n}-\varepsilon}\right). \label{gauss-mix}\end{equation}
At $\varepsilon=0$ it coincides with (\ref{sq-vac}). In the {\it thermal\/}
state we have $\varepsilon=\bar{n}$, $\rho_n=\bar {n}^n/\left(1+\bar {n}
\right)^{n+1}$, and $\gamma_{therm} =\gamma_0\exp(\chi\bar{n})$.
The last expression holds provided that $\bar{n}\ll (Q\ln{Q})^{-1/2}$. Note
that this restriction does not forbid the inequality $\chi\bar{n}\gg 1$.
With the same value of $\bar{n}$, at $\chi\bar{n}\gg 1$ the tunneling rate
from the squeezed (pure) vacuum state turns out much greater than that from
the thermal one. However, the thermal state decays faster than the coherent
one under the same conditions. These examples show that the decay rates are
very sensitive to the details of the energy distribution in the wave packet,
so it is difficult to find a general law.

An example of a Gaussian packet with nonzero means of the quadratures is the
mixture of the coherent and thermal states \cite{Sem,DOM,Moll}, when the
``shifted Planck distribution function'' is expressed in terms of the
Laguerre polynomials:
\begin{equation} \rho_n=\frac{n_{th}^n}{\left(1+n_{th}\right)^{n+1}}
\exp\left[-\frac {|\alpha|^2}{1+n_{th}}\right]L_n\left(-\frac{|\alpha|^2}
{n_{th}(1+n_{th})}\right).\label{shift}\end{equation}
In this case sum (\ref{main}) can be calculated exactly with the aid of
Eq. 10.12(18) from \cite{Bateman}. For $n_{th}\ll 1$ and $|\alpha|^2\ll 1$
the total decay rate equals the product of the coherent and thermal
decay rates:
\begin{equation} \gamma_{shift}(|\alpha|,n_{th})=\gamma_0\exp\left(\chi
n_{th}\right) I_0\left(2|\alpha|\sqrt{\chi}\right)=\gamma_{therm}\gamma_{coh}.
\label{tun-shift}\end{equation}

An example of a nonGaussian wave packet is the {\it even coherent state\/}
introduced in \cite{even},
\[\mid\alpha;+\rangle=
\left\{2\left[1+\exp\left(-2|\alpha|^2\right)\right]\right\}^{-1/2}
\left(\mid \alpha\rangle + \mid -\alpha\rangle\right),\]
with the quantum distribution function
\[ \rho_{2n}^{(+)}=\frac{|\alpha|^{4n}}{(2n)!\cosh|\alpha|^2}, \quad
\rho_{2n+1}^{(+)}=0, \qquad \bar{n}^{(+)}=|\alpha|^2\tanh|\alpha|^2\,.\]
In this case the total decay rate is proportional to the sum of the usual
and the modified Bessel functions of the argument $2|\alpha|\sqrt{\chi}$,
provided that $|\alpha|^2\ll (\mu_{\nu}\ln Q)^{-1}$. Then $\bar{n}=
|\alpha|^4$, and $\sigma_n=2\bar{n}$ at $|\alpha|\ll 1$. Therefore
\[\gamma_+=\frac12\gamma_0\left[I_0\left(2\bar{n}^{1/4}\sqrt{\chi}\right)+
J_0\left(2\bar{n}^{1/4}\sqrt{\chi}\right)\right]. \]
We see that the even coherent state is less stable with respect to tunneling
than the Glauber coherent state with the same value of $\bar{n}\ll 1$.
For all distributions, $\gamma\approx\gamma_0$ if $\chi\bar{ n}\ll 1$.

\section{Decay of slightly deformed excited states}
An {\it odd coherent state\/} \cite{even}
\[\mid\alpha;-\rangle=
\left\{2\left[1-\exp\left(-2|\alpha|^2\right)\right]\right\}^{-1/2}
\left(\mid \alpha\rangle - \mid -\alpha\rangle\right),\]
\[ \rho_{2n+1}^{(-)}=\frac{|\alpha|^{4n+2}}{(2n+1)!\sinh|\alpha|^2}, \quad
\rho_{2n}^{(-)}=0, \qquad \bar{n}^{(-)}=|\alpha|^2\coth|\alpha|^2\]
is an example of the deformed first excited oscillator state at
$|\alpha|\ll 1$. Its decay rate equals
\[\gamma_-=\frac{\gamma_1}{2\chi|\alpha|^2}\left[I_0\left(2|\alpha|
\sqrt{\chi}\right)-J_0\left(2|\alpha|\sqrt{\chi}\right)\right], \qquad
\gamma_1=\gamma_0\chi. \]
The limitations on $|\alpha|$ are the same as above, but now $|\alpha|^4
=3\left(\bar{n}-1\right)$.

Another example is the {\it odd squeezed state\/} \cite{even}:
\[\mid z;-\rangle =\exp\left[\frac z2 \left(\hat{a}^{\dag}\right)^2\right]
\mid 1\rangle, \qquad \rho_{2n+1}^{(z)}=\frac{(2n+1)!}{(n!)^2}\left|\frac
z2\right|^{2n}, \quad \rho_{2n}^{(z)}=0.\]
The expression for the total decay rate is similar to formula (\ref{sq-vac})
for the squeezed vacuum state:
\[ \gamma_z=\gamma_1 I_0(\chi|z|),\]
provided that $|z|^2=(\bar{n}-1)/3\ll( Q\ln Q)^{-1}$.

It is not difficult to perform the calculations also for two families of
deformed $m$-quantum states. The first one corresponds to the {\it
photon-added coherent states\/} (PACS) \cite{Ag-Tar}:
\[\mid\alpha,m\rangle =\left(\hat {a}^{\dag}\right)^m\mid\alpha\rangle,
\qquad \rho_n^{pacs}=\frac{n!|\alpha|^{2(n-m)}}{m![(n-m)!]^2},\quad n\ge m \]
(we assume that $|\alpha|\ll 1$). Then the total decay rate is given by the
formula similar to (\ref{gam-coh}),
\[ \gamma_{pacs}=\gamma_m I_0\left(2|\alpha|\sqrt{\chi}\right), \qquad
|\alpha|^2 \ll 1/\ln{Q},\]
but with another meaning of parameter $|\alpha|$, since now
$ \bar{n}-m=(m+1)|\alpha|^2=\sigma_n $.

The second family consists of the {\it displaced number states\/}
\cite{Boit,Roy,Venk,Oliv,Wun} $\mid m,\alpha\rangle =\exp\left(\alpha\hat
{a}^{\dag}-\alpha^*\hat{a} \right)\mid m\rangle$, whose quantum distribution
function is expressed in terms of the associated Laguerre polynomials:
\[ \rho_n^{disp}=\frac{n!}{m!}\left[|\alpha|^{(n-m)}L_n^{(m-n)}\left(
|\alpha|^2\right)\right]^2\exp\left(-|\alpha|^2\right). \]
In this case, using the identity 10.12(19) from \cite{Bateman}
\[ \sum_{n=0}^{\infty} z^n L_n^{(\alpha-n)}(x)=e^{-zx}(1+z)^{\alpha} \]
and applying the same approach that led to Eq. (\ref{gam-squeez}), one can
obtain the formula
\[ \sum_{n=0}^{\infty} y^n\left[ L_n^{(m-n)}(x)\right]^2=\frac1{2\pi}\int_0
^{2\pi}d\varphi\exp\left(-2x\sqrt{y}\cos\varphi\right)\left(1+y+2\sqrt{y}
\cos\varphi\right)^m .\]
Consequently, the total decay rate can be expressed as some combination of
the modified Bessel functions of $2|\alpha|\sqrt{\chi}$  with different
integer indices. However, in the most interesting case, when $2|\alpha|
\sqrt{\chi}\gg 1$, the steepest descent method leads to a simple formula
\[ \gamma_{disp}=\gamma_m\left(4\pi|\alpha|\sqrt{\chi}\right)^{-1/2}
\exp\left(2|\alpha|\sqrt{\chi}\right), \quad
|\alpha|^2=\bar{n}-m \ll 1/\ln{Q}, \quad \sigma_n=(2n+1)|\alpha|^2.\]

\section{Conclusion}
Two new results seem to be the most important. Firstly, we have found the
universal Poisson distribution of the partial decay rates from the energy
eigenstates in the parabolic potential well for a wide class of potential
barriers. Secondly, we have demonstrated that the tunneling decay rates
are very sensitive to the shape of the wave packet. In particular,
if one has initially not an exact energy eigenstate, but a combination
(pure or mixed) of the states with different energies, then the decay rates
may be quite different, even when the average energy, coordinate
and momentum variances, etc., are almost the same.
This fact may be important for the analysis of various phenomena related to
the tunnel effect, when the initial state is not known absolutely exactly.

\section*{Acknowledgement}
This research was partially supported by Russian Basic Research Foundation.

\end{document}